\documentclass[psfig]{aa}

\input psfig.sty

\newcommand{\gsim}{\raisebox{-3.8pt}{$\;\stackrel{\textstyle >}{\sim}\;$}}
\newcommand{\lsim}{\raisebox{-3.8pt}{$\;\stackrel{\textstyle <}{\sim}\;$}}

\newcommand{\Msol}{$M_{\odot}$}

\newcommand{\etal}{\mbox{{\rm et~al.\ }}}

\newcommand{\Oxygen}{\mbox{${\rm {\rm ^{16}O}}$}}

\newcommand{\Sulfur}{\mbox{${\rm {\rm ^{32}S}}$}}

\def\smallskip{\vskip 6pt}

\begin{document}

\thesaurus{       }

\title {On star formation and chemical evolution\\
in the Galactic disc} 

\author{L.\ Portinari and C.\ Chiosi }

\institute{
Department of Astronomy, University of Padova, Vicolo dell'Osservatorio 5, 
35122 Padova, Italy (portinari, chiosi@pd.astro.it)}

\offprints{L.\ Portinari}
\date{Received 10 June 1999 / Accepted 27 August 1999}

\maketitle
\markboth{Portinari and Chiosi: On star formation and chemical 
evolution in the Galactic disc}{}


\begin{abstract}

The abundance gradients and the radial gas profile of the Galactic disc
are analysed by means of a model for the chemical evolution of galaxies.
As one of the major uncertainties in models for galaxy evolution is the
star formation (SF) process, various SF laws are considered,
to assess the response of model predictions to the different assumptions. 
Only some SF laws are successful in reproducing the metallicity gradient, 
and only if combined with a suitable infall timescale increasing outward
(inside--out formation scenario).
Still, it is difficult to reproduce at the same time also the observed 
gas distribution; we therefore suggest further improvements for
the models.

\end{abstract}


\section{Introduction}

Modelling and understanding the formation and evolution of galaxies is a
major issue of modern astrophysics, involving a variety of approaches: 
dynamical and hydro-dynamical simulations, chemical evolution models,
stellar population synthesis techniques.
Any galaxy model requires a recipe for the star formation (SF) process, 
which unfortunately is still poorly known, especially on the large scales 
relevant to galaxy evolution.
A variety of SF laws has been suggested in literature especially for spiral
galaxies, where we can observe on--going SF on large scales. In chemical models
for galactic discs, different prescriptions for the SF process are used by 
different authors. This is true also for models of the disc of the Milky Way,
a system which we can observe and study in much closer detail than any other.
It is often hard to compare models and conclusions by different authors,
since it is not clear to what extent model results have a general validity
or rather depend on differing assumptions for the SF law.

The present work is addressed at testing different prescriptions for the SF law
in the Galactic disc. 
To this aim we use a model for the chemical evolution of galaxies, able to 
handle different options for the SF law (Section~\ref{modelSFR}). 
We review extant data on the abundance gradients and on the radial gas and star
formation rate (SFR)  
profiles of the Galactic disc (Section~\ref{data}). To these observational
counterparts we compare models with the various SF laws,
trying to discriminate, if possible, a ``best--fit'' law 
(Section~\ref{gradmodelling}). Summary and conclusions are finally drawn in 
Section~\ref{conclusions}.

A similar analysis was already performed by Prantzos \& Aubert (1995), but an 
upgrade of their study is suggested since some basic observational results
have meanwhile changed, especially regarding the metallicity gradient
(\S~\ref{data}).


\section{The chemical model and the SF laws}
\label{modelSFR}

The chemical evolution of the Galactic disc is simulated by means of the
model of Portinari \etal (1998, hereinafter PCB98), to which we refer for 
a detailed description. Here we just overview the main features of the model.

It is an open model with continuous infall, as suggested by the metallicity
distribution of G--dwarfs in the Solar Vicinity (Lynden-Bell 1975) and by 
dynamical simulations of disc formation (e.g.\ Larson 1976). Open models 
also seem to be appropriate for galaxies more in general --- for instance,
see Bressan \etal (1994) for the analogous of the G--dwarf problem in 
elliptical galaxies. The infalling gas, assumed to be primordial, settles 
onto the disc at an exponentially decreasing rate with timescale $\tau(r)$:
\[ \dot{\sigma}_{inf}(r,t) = A(r) \, e^{-\frac{t}{\tau(r)}} \]
$A(r)$ is fixed by the assigned final (present--day) surface density profile of
the model disc, an exponential with scale--legth $r_d$:
\[ \sigma(r,t_G) = \sigma_{\odot} \, e^{ - \frac{r-r_{\odot}}{r_d}} \]
where $\sigma_{\odot} = \sigma(r_{\odot},t_G)$, total surface density at the
Solar radius at the final Galactic age $t_G$.

In PCB98 the chemical model was applied only to the Solar Neighbourhood, while
in the present paper several concentric cylindrical shells (rings) are 
calculated, spanning the Galactic disc from 2 to 20 kpc.
Each ring, typically 2 kpc wide, consists of a homogenous mixture 
of gas and stars which is assumed to evolve independently, neglecting any 
possible radial flows. The instantaneous recycling approximation (IRA) is
relaxed in the calculation of stellar ejecta.
The chemical evolution of the interstellar medium (ISM) in each ring $r$ 
is described by the set of equations:
\begin{equation}
\label{dGi/dt}
\begin{array}{l l}
\frac{d}{dt} G_i(r,t) = & - X_i(r,t) \, \Psi(r,t) \,+ \\
 & +\, \int_{M_l}^{M_u} \Psi(r,t-\tau_M) \, R_i(M) \, \Phi(M) \, dM \,+ \\
 & +\, \left[\frac{d}{dt} G_i(r,t) \right]_{inf}
\end{array}
\end{equation}
where the $G_i$'s are the surface gas densities of each chemical species $i$,
normalized to the total surface density at the present age of the Galaxy 
$\sigma(r,t_G)$. The first term on the right-hand side represents the depletion
of species $i$ from the gaseous phase due to SF; the second term is 
the amount of species $i$ ejected back to the ISM by dying stars; the third 
term is the contribution of the infalling gas. 

The various ingredients of Eq.~(\ref{dGi/dt}) and its numerical solution 
are described in detail in PCB98. With respect to that model, the only 
difference is that here we adopt different 
options for the SFR, $\Psi(r,t)$. We describe here below the various SF laws
considered, together with their physical underpinnings.


\subsection{Schmidt law}

A few analytical prescriptions for the SF law in galactic discs are available
in literature. The easiest physical law is Schmidt (1959) law:
\[ SFR \, \propto \, \sigma_g^{\kappa} \]
where $\sigma_g$ is the surface gas density.
In spite of its crudeness, Schmidt law still remains, after 40 years, 
the most popular SF law in galaxy models. 

Recent empirical determinations in spiral and starburst
galaxies indicate a ``Schmidt exponent'' $\kappa \sim 1.5$ 
(Kennicutt 1998). The observed correlation is quite good between 
the SFR and the total density of cold gas (HI+H$_2$), while no clear 
correlation apparently holds between the SFR and the sole H$_2$ component ---
at odds with expectations since stars form within molecular clouds.
Therefore, we consider here a Schmidt law with exponent 1.5, involving the 
overall gas density with no distinction between the atomic and the molecular 
phase.

In the formalism of our model, we translate Schmidt law into a normalized
SFR per unit surface density (the function $\Psi$ in Eq.~\ref{dGi/dt}) as:
\begin{equation}
\label{SFSchmidt}
\Psi(r,t) = - \left[ \frac{d G}{dt} 
\right]_* (r,t) = \nu \, \left[ \frac{\sigma(r,t_G)}{\sigma(r_{\odot},t_G)} 
\right]^{\kappa-1} \, G^{\kappa}(r,t)
\end{equation}
where we have introduced the constant normalization factor 
$\sigma(r_{\odot},t_G)^{-(\kappa-1)}$ for the sake of expressing the SF
efficiency $\nu$ in [$t^{-1}$]. We adopt $\kappa=1.5$.

Hereinafter, chemical models adopting the Schmidt law (\ref{SFSchmidt}) will be
labelled with {\sf S15}.

\bigskip
Other SF laws still assume that the SFR depends on the gas density, but also 
take into account additional phenomena affecting the SF process. As a result,
the efficiency with which the available gas is turned to stars depends
on the galactocentric distance $r$. Such laws can be written as 
``modified'' Schmidt laws, 
with a proportionality coefficient, or efficiency, 
varying radially over the disc: 
\[ SFR = \nu(r) \, \sigma_g^{\kappa} \]
The detailed radial behaviour is related to the physical 
processes involved, as we discuss in the following
cases.


\subsection{Spiral density waves trigger}

According to a theory dating back to Roberts (1969) and Shu \etal (1972), SF
takes place in spiral arms due to gas compression in density 
waves. In this framework, Oort (1974) suggested that the SFR at a given
galactocentric distance is determined by the rate at which
the orbiting gas crosses spiral arms:
\[ SFR \propto (\Omega(r)-\Omega_p) \, \sigma_g^{\kappa} \]
where $\Omega(r)$ is the angular rotation velocity as a function of radius 
and $\Omega_p$ is the angular velocity of the spiral pattern.
In the scenario for disc galaxy formation within a pre-existing dark halo, and
also according to viscous disc models (e.g.\ Saio \& Yoshii 1990), the 
rotation curve is time--independent and keeps the same shape throughout 
the evolution. For a flat rotation curve typical of galactic discs 
$\Omega(r) \propto r^{-1}$, and since usually $\Omega_p \ll \Omega(r)$,
we roughly get:
\[ SFR \propto r^{-1} \, \sigma_g^{\kappa} \]
Such a SF law was adopted, for instance, by 
Wyse \& Silk (1989) and by Prantzos \& Silk (1998).

A similar radial dependence of the SF efficiency
was also obtained by Wang \& Silk (1994), who treated the SF process as a 
consequence of gravitational instabilities in the disc. These are related to 
the local Toomre's stability parameter $Q$, itself related to the local 
epyciclic frequency 
\[ \kappa_e(r) \propto \Omega(r) \propto r^{-1} \]

Normalizing the expression so that the efficiency coefficient $\nu$ is in 
units of $[t^{-1}]$, in our chemical model we express this ``Oort SF law'' as:
\begin{equation}
\label{SFWyse}
\Psi(r,t) = \nu \left[\frac{r}{r_{\odot}} \right]^{-1} \, \left[ 
\frac{\sigma(r,t_G)}{\sigma(r_{\odot},t_G)} \right]^{\kappa-1} G^{\kappa}(r,t)
\end{equation}

Kennicutt (1998) noticed that his empirical relation between the current
SFR and the 
gas surface density, yielding $\kappa \sim 1.5$ if fitted with a Schmidt
law, is equally well fitted with an Oort--type law with $\kappa = 1$. 
Therefore, in the Oort SF~law~(\ref{SFWyse}) we will adopt an exponent 
$\kappa=1.0$. 
Such models will be labelled hereinafter with {\sf O10}.

For the sake of comparison with other authors, we will though consider also
the case of an Oort--type law with $\kappa=1.5$ (models {\sf O15}, see 
\S~\ref{modelsO15}).


\subsection{Self-regulating gravitational settling + feed-back}
\label{self-regSF}

Other prescriptions describe SF as a self-regulation process between the
gravitational settling of the gas onto the plane of the disc and the energy 
feed-back from massive stars. The former phenomenon leads to compression and 
cooling of the gas layer, favouring its collapse into clouds and stars. 
Stellar winds and supernova explosions from massive stars provide a heating 
and turbulence source that supports the gas layer against further compression 
and cooling. The SFR results from the balance between these two opposing 
effects.
Where the total surface mass density $\sigma$ is larger, gravitational 
settling and the related cooling are more efficient, and a larger SFR is
required to reach the equilibrium feed-back.
As a result, the SFR  depends both on the gas surface density $\sigma_g$ and 
on $\sigma$.
Since in spiral discs $\sigma$ displays an exponential profile,
the SF efficiency inherits an exponential radial decline.
Such a self-regulating SF process was first proposed by Talbot \& Arnett 
(1975) as:
\begin{equation}
\label{propSFTA}
SFR \, \propto \, \sigma(r)^{\kappa-1} \, \sigma_g^{\kappa}
\end{equation}
A similar formulation was suggested later by Dopita (1985).
Self-regulating SFRs depending on the local surface mass density are also 
obtained in full chemo-dynamical models (Burkert \etal 1992).

More recently, Dopita \& Ryder (1994) discovered that an empirical relation
holds in outer spiral discs between the H$_{\alpha}$ emission, tracing
on--going SF, and the surface I--band brightness, tracing the underlying
old stellar component and thus the total surface density. This relation 
provides the observational counterpart to a SFR depending on $\sigma$.
Dopita \& Ryder derived a theoretical SF law
(strongly reminiscent of the one by Talbot \& Arnett):
\begin{equation}
\label{propSFDR}
SFR \propto \sigma^n \, \sigma_g^m
\end{equation}
with $n=1/3$, $m=5/3$, which fits very well with the
observed relation. It can also account for the observed correlation between
oxygen abundance and surface brightness in spiral discs (Ryder 1995). 

In the formalism of our chemical model,
normalizing the SFR~(\ref{propSFDR}) so that the efficiency coefficient $\nu$
is always in units of $[t^{-1}]$, we write:
\begin{equation}
\label{SFDopita}
\Psi(r,t) = \nu \left[ \frac {\sigma^n(r,t) \, 
\sigma^{m-1}(r,t_G)}
{\sigma(r_{\odot},t_G)^{n+m-1}} \right] \, G^m(r,t)
\end{equation}
and we adopt $n=1/3$, $m=5/3$. Chemical models with this SF law will be 
labelled with {\sf DR}.

In PCB98 the SF law by Talbot \& Arnett (Eq.~\ref{propSFTA}) was adopted. 
That SF law, with Schmidt--like exponent {\mbox {$\kappa=1.5$}}, naturally 
leads to 
similar predictions as models {\sf DR}, since the two SF laws are analogous.
Therefore, it's not worth in the following discussion to present also
models with Talbot \& Arnett's law separately.


\subsection{The Initial Mass Function}
\label{IMFlaw}

Besides the infall timescale $\tau(r)$ and the SF efficiency $\nu$ of the
various SF laws, another important model parameter is related to the
Initial Mass Function (IMF). We adopt a power--law IMF:
\[ \Phi(M) \, \propto \, M^{-\mu} \]
where $\mu$=1.35 for $M<2$~\Msol\ as in Salpeter (1955) and $\mu$=1.7
for $M>2$~\Msol\ following Scalo (1986). 
Since the bulk of chemical enrichment is due to stars with 
{\mbox{$M \ge 1$~\Msol,}} a meaningful parameter is the fraction $\zeta$ 
of the total stellar mass distributed in stars above 1~\Msol. 
%
\[ \zeta\,=\,\int_{1}^{100}\Phi(M)\,dM \]
Fixing the ``scaling fraction'' $\zeta$ is equivalent to fixing the lower end
$M_l$ of the IMF (see PCB98 and Chiosi \& Matteucci 1982 for further details).
In our models we always keep the IMF constant all over the disc and in time.

\begin{table*}[t]
\caption{Observed abundance gradients of various elements. $\Delta r$ is the
galactocentric radial range covered by each resective study (in kpc). 
Gradients are expressed in dex/kpc.}
\label{gradientstab}

\begin{center}
\begin{tabular}{l l|c|c c c c}
\hline
tracer & reference & $\Delta r $ 	& $\frac{d[O/H]}{dr}$
				& $\frac{d[N/H]}{dr}$
				& $\frac{d[S/H]}{dr}$ \\
\hline
HII regions & Shaver \etal (1983)$^{(1)}$ & 4--13 & -0.07$\pm$0.015
                                & -0.09$\pm$0.015 
				& -0.01$\pm$0.02 \\
(optical) & Fich \& Silkey (1991) & 12--18 & ---  
	                        & $\sim$ 0
				& --- \\
       & Vilchez \& Esteban (1996) & 12--18 & -0.036$\pm$0.02 
				& -0.009$\pm$0.020 
				& -0.041$\pm$0.020 \\
\hline
HII regions & Rudolph \etal (1997)$^{(2)}$ & 0--17 & -0.079$\pm$0.009 
				& -0.111$\pm$0.012 
				& -0.079$\pm$-0.009 \\
(FIR)  & Afflerbach \etal (1997) & 0--12 & -0.064$\pm$0.009 
			        & -0.072$\pm$0.006 
				& -0.063$\pm$0.006 \\
\hline
OB stars & Smartt \& Rolleston (1997) & 6--18 & -0.07$\pm$0.01
				   &  --- 
				   &  --- \\
         & Gummersbach \etal (1998)   & 5--14 & -0.07$\pm$0.02
				   &  -0.08$\pm$0.02
				   &  --- \\
\hline
Type II PN\ae\ & Maciel \& K\"oppen (1994) & 4--14 & -0.06$\pm$0.01
				   &  ---
				   &  -0.07$\pm$0.01 \\
\hline
\end{tabular}
\end{center}

{\scriptsize $^{(1)}$ Rescaled to $r_{\odot}=8.5$~ kpc.\\
$^{(2)}$ Includes the data of Simpson \etal (1995).}
\end{table*}


\section{Observational constraints}
\label{data}

In this section we review the available observational data on the
abundance gradients and on the gas and SFR radial profiles in the Galactic
disc. In the next section we will discuss
how predictions from chemical models, adopting the different SF laws,
compare to these observational data. 


\subsection{Metallicity gradients}

Negative radial gradients of metallicity in spiral discs have long been 
established. Abundance data are usually derived from 
HII regions and from bright blue stars.
These young objects mainly trace the present-day gradient for oxygen, and then
for nitrogen, sulfur, neon and argon.
For the Milky Way, the first extensive study of abundances in HII regions
was presented by Shaver \etal (1983). Later studies based on optical spectra, 
in the inner and outer Galaxy, are by Fich \& Silkey (1991) and Vilchez \& 
Esteban (1996). 
Abundances in  HII regions have been also measured from FIR lines (Simpson 
\etal 1995, Afflerbach \etal 1997, Rudolph \etal 1997).
Studies on HII regions generally agree on the existence of
a Galactic oxygen gradient around --0.07 dex/kpc.
Maciel \& K\"oppen (1994) used Type II planetary nebul\ae\
as tracers of the present-day gradient, which they derive to be --0.06 dex/kpc,
consistently with studies on HII regions.

At odds with nebular studies, 
the stellar spectra
of OB stars and associations seemed to suggest a very flat oxygen gradient
(Gehren \etal 1985; Fitzsimmons \etal 1990, 1992; Kaufer \etal 1994; 
Kilian-Montenbruck \etal 1994).
The discrepancy disappears in the recent, extensive studies by Smartt \& 
Rolleston (1997) 
and by Gummersbach \etal (1998), both deriving a gradient of --0.07 dex/kpc.
This revision of the oxygen gradient in our Galaxy, in fact, suggested to us
to revise as well the study by Prantzos \& Aubert (1995) on the SF and chemical
evolution of the Galactic disc.

\begin{figure*}[ht]
\centerline{\psfig{file=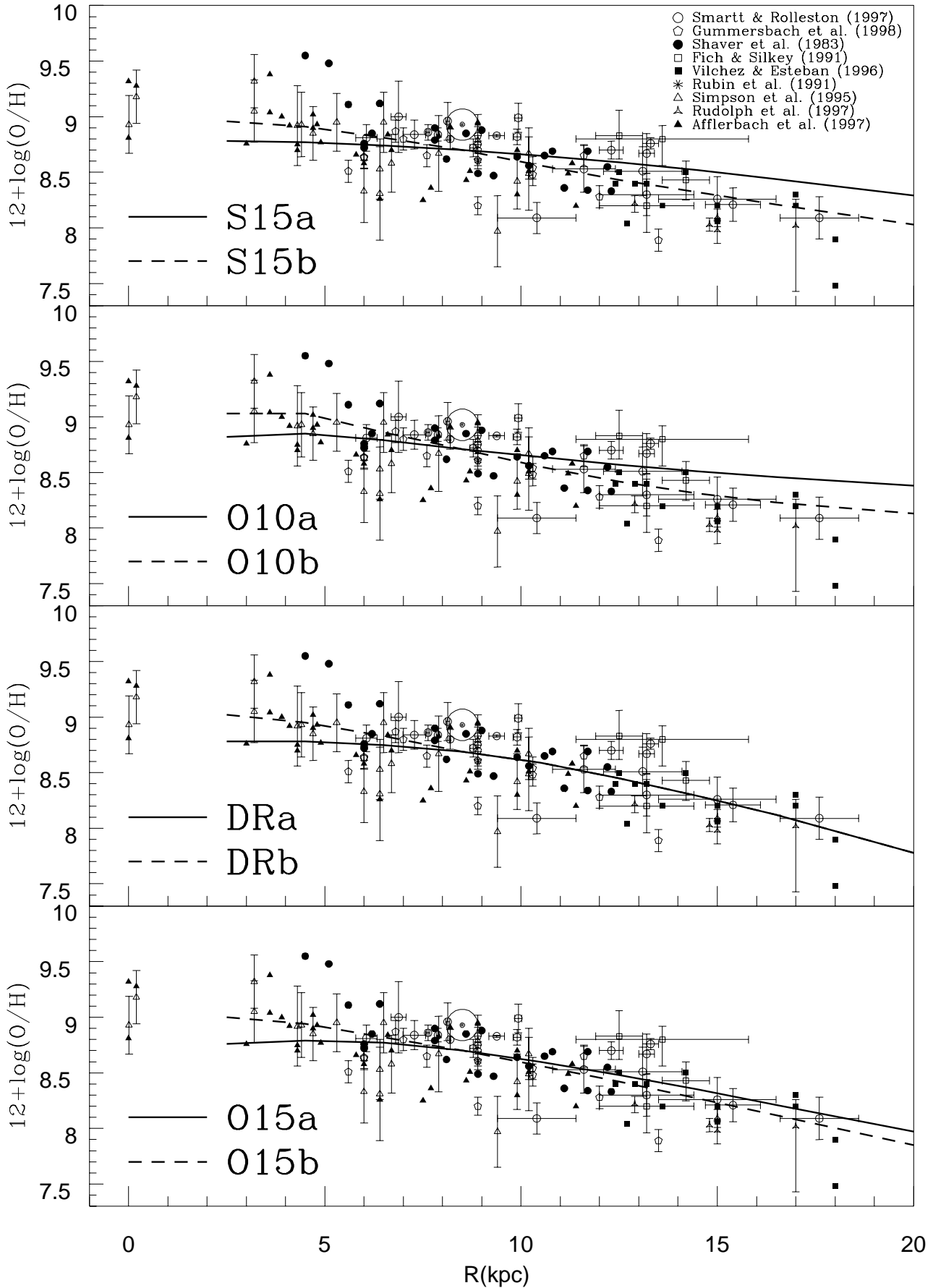,width=18truecm}}
\caption{Predicted radial oxygen gradient from the various models (see text)
compared to observational data. The location of the Sun $\odot$ is also
displayed for comparison.}
\label{gradOH}
\end{figure*}

In spite of the overall agreement, it is still controversial whether
the slope of the oxygen gradient is constant all over the disc or not. A 
flattening in the outer zones 
has been claimed by Vilchez \& Esteban (1996) and Fich \& Silkey (1991), 
while other studies do not suggest such a trend (Smartt \& Rolleston 1997; 
Rudolph \etal 1997; Afflerbach \etal 1997). In the overall, considering all 
the different data and tracers together (Fig.~\ref{gradOH}), we consider 
that a single--slope gradient is a good enough description, and take
\[ \frac{d[O/H]}{dr} = -0.06 \, \div \, -0.07 \, {\rm dex} \, {\rm kpc^{-1}} \]
as our estimate of the oxygen gradient throughout the Galactic disc.

A similar controversy also exists for the [N/H] gradient 
(Fig~\ref{gradNSFeHfig}, top panel) which flattens out
according to optical studies (Shaver \etal 1983; Fich \& Silkey 1991; 
Vilchez \& Esteban 1996), while it remains roughly constant according to FIR 
measurements (Simpson \etal 1995; Rudolph \etal 1997; Afflerbach \etal 1997).
The problem seems to arise from the very different,  by {\mbox {0.3--0.4~dex}},
nitrogen abundance determinations around $r \sim r_{\odot}$ in the two sets 
of analysis (Rudolph \etal 1997).

The very flat gradient of sulfur found by Shaver \etal (1983) is not confirmed
in later studies; the sulfur gradient (Fig.~\ref{gradNSFeHfig}, middle panel)
seems rather to follow the oxygen gradient,
or equivalently the [S/O] ratio seems to be constant with radius, as
expected for an abundance ratio of two primary elements produced by the same
source (type II SN\ae). There seems to be no longer need to account for a 
gradient in the [S/O] ratio by invoking slightly different progenitors 
for the two elements, like SN\ae\ from massive stars of different mass ranges 
(Matteucci \& Fran\c{c}ois 1989).

The abundance gradients as determined in the above mentioned
studies are listed in Table~\ref{gradientstab}; the corresponding data are 
displayed in Fig.~\ref{gradOH}. 
For the sake of completeness, from the above mentioned studies we plot also 
the data concerning the Galactic Centre. However, we will not 
consider these data as a constraint for chemical models of the disc, since the
stellar population of the Galactic Centre is likely to have evolved on its own,
rather than represent an extension of the 
disc population to $r \rightarrow 0$. Our disc models will consider only
regions down to about $r=2$~kpc, as further inward 
the Bulge becomes the dominating Galactic component.

The radial gradient of [Fe/H] is mainly traced by open clusters. An up-to-date
determination based on a homogeneous sample of open clusters can be
found in Carraro \etal (1998), who find 
\[ \frac{d[Fe/H]}{dr} \sim -0.07 \div -0.09 \, {\rm dex} \, {\rm kpc}^{-1} \]
in good agreement with previous results 
(Janes 1979; Friel 1995).

\begin{table*}[t]
\caption{Parameter values and resulting metallicity gradients of the models.}
\label{modelstab}
\medskip
\begin{center}
\begin{tabular}{l|c|c|c|l|c l|c l}
\hline
 & & & & & & & \\
 SF law & $\nu$ & $\zeta$ & $r_d$ [kpc] & model & \multicolumn{2}{|c|}{$\tau(r)$ [Gyr]} & \multicolumn{2}{|c}{$\frac{d[O/H]}{dr}$ [dex kpc$^{-1}$]} \\
 & & & & & & & \\
\hline
           &      &     &        & {\sf S15a}  & 3                                           & (uniform)           & --0.03  &   \\
{\sf S15}  & 0.35 & 0.2 & 4 & & & & & \\
           &      &  &  & {\sf S15b}  & $3 \, e^{\frac{r-r_{\odot}}{r_d}}$ &                      & --0.055 &            \\
 & & & & & \multicolumn{2}{|p{6.5truecm}|}{(from $\tau \sim 0.7$ at $r=2.5$ to $\tau \sim 50$ at $r=20$)} & & \\
\hline
           &      &  &  & {\sf O10a} & 3                            & (uniform)           & --0.03  &                   \\
{\sf O10}  & 0.19 & 0.2 & 4 & & & & \\
	   &	  &  &  & {\sf O10b} & $3 \, e^{\frac{r-r_{\odot}}{r_d}}$ &                      & --0.06  &             \\
 & & & & & \multicolumn{2}{|p{6.5truecm}|}{(from $\tau \sim 0.7$ at $r=2.5$ to $\tau \sim 50$ at $r=20$)} & & \\
\hline
           &      &  &  & {\sf DRa}   & 3                            & (uniform)           & --0.07 & ($r > r_{\odot}$) \\
           &      &  &  &             &                              &                      & flat   & ($r < r_{\odot}$) \\
 & & & & & & & \\
	   &	  &  &	& {\sf DRb}   & 3                            & ($r \geq r_{\odot}$) & --0.07  &                  \\
{\sf DR}   & 0.45 & 0.2 &  4  &        & $\sim 0.1+ \frac{r-2.5}{2}$ & ($r < r_{\odot}$)    &        &                   \\
 & & & & & \multicolumn{2}{|p{6.5truecm}|}{(from $\tau \sim 0.1$ at $r=2.5$ to $\tau=3$ at $r \geq r_{\odot}$)} & & \\
 & & & & & & & & \\ 
	   &	  &  &	& {\sf DRc}   & 3                            & ($r \geq 4.5$~kpc) & --0.06  &                   \\
           &      &  &  &             & $\sim 1+ (r-2.5)$            & ($r < 4.5$~kpc)    &        &                   \\
 & & & & & \multicolumn{2}{|p{6.5truecm}|}{(from $\tau \sim 1$ at $r=2.5$ to $\tau=3$ at $r \geq 4.5$)} & & \\
\hline
           &      &  &  & {\sf O15a} & 3                            & (uniform)           & -0.05  & ($r > r_{\odot}$) \\
           &      &   &  &             &                              &                      & flat   & ($r < r_{\odot}$) \\
{\sf O15} & 0.35 & 0.2 & 4 & & & & \\
	   &	  &  &	& {\sf O15b} & $\sim 3 \frac{r}{r_{\odot}}$ &                      & -0.065 &                   \\
 & & & & & \multicolumn{2}{|p{6.5truecm}|}{(from $\tau \sim 1$ at $r=2.5$ to $\tau \sim 7$ at $r=20$)} & & \\
\hline
\hline
           &     &  &  & {\sf DR$\tau$9a}   & 9                            & (uniform)  & --0.06  & ($r \gsim 12$~kpc) \\
           &      &  &  &             &                              &                      & flat  & ($r \lsim 12$~kpc) \\
{\sf DR$\tau$9}   & 0.95 & 0.25  & 4 & & & & \\
   &	  &  &	& {\sf DR$\tau$9b}   & $9 \, e^{\frac{r-r_{\odot}}{3~kpc}}$             & ($r \leq 10$~kpc)   & --0.06 & \\
           &      &     &      &      & $\sim 50$ & ($r > 10$~kpc)    &        &                   \\
\hline
\hline
           &      &  &  & {\sf S15$r_d$35a}  & 3                                           & (uniform)           & --0.035  &   \\
{\sf S15$r_d$35}  & 0.35 & 0.2 & 3.5 & & & & & \\
           &      &  &  & {\sf S15$r_d$35b}  & $3 \, e^{\frac{r-r_{\odot}}{r_d}}$ &                      & --0.06 &            \\
 & & & & & \multicolumn{2}{|p{6.5truecm}|}{(from $\tau \sim 0.7$ at $r=2.5$ to $\tau \sim 50$ at $r=20$)} & & \\
\hline
           &      &  &  & {\sf O10$r_d$35a} & 3                            & (uniform)           & --0.03  &                   \\
{\sf O10$r_d$35}  & 0.19 & 0.2 & 3.5 & & & & & \\
	   &	  &  &	& {\sf O10$r_d$35b} & $3 \, e^{\frac{r-r_{\odot}}{r_d}}$ &                      & --0.06  &             \\
 & & & & & \multicolumn{2}{|p{6.5truecm}|}{(from $\tau \sim 0.7$ at $r=2.5$ to $\tau \sim 50$ at $r=20$)} & & \\
\hline
           &      &  &  & {\sf DR$r_d$35a}   & 3                            & (uniform)           & --0.075 & ($r > r_{\odot}$) \\
           &      &   &  &            &                              &                      & flat   & ($r < r_{\odot}$) \\
{\sf DR$r_d$35} & 0.45 & 0.2 & 3.5 & & & & & \\
	   &	  &  &	& {\sf D$r_d$35Rb}   & 3                            & ($r \geq 6.5$) & --0.075  &                  \\
& & &          &  & $\sim 1 + \frac{r-2.5}{2}$ & ($r < 6.5$)    &        &                   \\
 & & & & &\multicolumn{2}{|p{6.5truecm}|}{(from $\tau \sim 0.1$ at $r=2.5$ to $\tau=3$ at $r \geq r_{\odot}$)} & & \\
\hline
\end{tabular}
\end{center}
\end{table*}


\subsection{Gas and SFR radial profiles}

The radial profile of the cold gas distribution (sum of HI and H$_2$) is taken
from 
the review by Dame (1993; Fig.~\ref{radlGlSFRfig}, top panel). For $r > 7$~kpc,
the profile is shallow with a plateau or a mild decline outward, while around
4--7~kpc a pronounced peak is due to the molecular ring. For 
{\mbox{$r < 3$~kpc}}, 
there is a rapid drop in the gas density.

Estimates for the SFR over the disc come from the distribution of SN remnants
(Guibert \etal 1978), of pulsars (Lyne \etal 1985), of Lyman--continuum 
photons (G\"usten \& Mezger 1982) and of molecular gas
(Rana 1991 and references therein). Following Lacey \& Fall (1985), since these
observables cannot directly yield the absolute SFR without assumptions on the 
IMF etc., we just normalize them to their value at the Solar ring, and consider
them as tracers of the radial profile SFR($r$)/SFR($r_{\odot}$)
(Fig.~\ref{radlGlSFRfig}, bottom panel). Unsurprisingly, the observed 
SFR profile
follows that of the gas distribution, with a peak around $r=4$~kpc and a milder
decrease outward of the Solar ring.

\section{Model results}
\label{gradmodelling}

In this section we compare model results for the various SF laws to the
observed radial profile of the disc. 

The models are calibrated as follows.
The age of the Galaxy is $t_G$=15~Gyr. The final density profile of the 
disc $\sigma(r,t_G)$ is exponential with scale-length $r_d$
and with a surface mass density at the Solar ring 
{\mbox{($r_{\odot}=8.5$~kpc)}} of
$\sigma(r_{\odot},t_G) = 50$~\Msol~pc$^{-2}$ (PCB98 and references 
therein). In most models, {\mbox{$r_d=4$~kpc}} and the infall timescale 
at the Solar ring is $\tau(r_{\odot}) = 3$~Gyr; different values for the
infall timescale at the Solar ring and for the scale--length are
discussed in~~\S~\ref{changetau} and~~\S~\ref{changerd}.
The adopted~~IMF~~is from 
\clearpage
\noindent
Scalo (1986), uniform over the disc and constant in time, with an 
``amplitude'' factor for SN\ae~Ia $A=0.07$ (see \S~\ref{IMFlaw} and PCB98 
for further details).
For each distinct SF law, the SF efficiency $\nu$ and the ``IMF scaling 
fraction'' $\zeta$ are calibrated so that 
at the Solar ring the model recovers the observed present-day surface gas 
density and oxygen abundance in the ISM (8--9 \Msol\ pc$^{-2}$ and 
[O/H]$\sim 8.7$ dex, respectively). 
With respect to the latter constraint, notice that the local ISM 
is oxygen-poor with respect to the Sun (e.g.\ Meyer \etal 1998, and 
Fig.~\ref{gradOH}), 
at odds with expectations from usual chemical models where metallicity 
always increases and should be higher in the present-day ISM than in
the Sun, formed about 5~Gyr ago. Different 
explanations have been put forward to account for this effect (dust depletion,
orbital diffusion, SN II pollution at Sun formation, recent metal-poor 
infall...), but all of them are rather
{\it ad hoc} or uneasy to model properly. Here we will not deal with this
problem; following Prantzos \& Aubert (1995) we just concentrate on 
the radial metallicity gradient, irrespectively of what the true
zero-point abundance at the Solar ring actually is. Since the oxygen gradient
is measured in young objects, we prefer to compare alike with alike
and calibrate the model onto the \Oxygen\ abundance of the local ISM, rather 
than onto the solar one. Needless to say that models so calibrated will 
not reproduce the local age--metallicity relation and the solar 
metallicity; but the particular choice of the zero--point will not affect 
the radial behaviour of the models.

Once we have so fixed the same Solar point
for all the models, we analyse the radial behaviour in the various cases.
The calibrated values for $\nu$ and $\zeta$ for the various models presented
are given in Table~\ref{modelstab}, together with model
predictions for the oxygen gradient.

First of all we consider the constraint provided by the oxygen gradient as
the best tracer of the overall metallicity (Wheeler \etal 1989), 
since it is the most abundant metal and the best sampled element in 
tracers of the present-day gradient (\S~\ref{data}).
Besides, its production and theoretical yields are the best understood of all 
the metals (Prantzos 1998), so the
modelling of the abundance evolution of oxygen is quite reliable.


\bigskip 
\subsection{Models with Schmidt law {\sf (S15)}}

When a Schmidt law is adopted (Eq.~\ref{SFSchmidt} with $\kappa = 1.5$),
if no radial variation of model parameters is assumed, the predicted gradient
is far too flat (model {\sf S15a}, Fig.~\ref{gradOH}). This is a well known
result in literature (e.g.\ Lacey \& Fall 1985; Edmunds \& Greenhow 1995).
Some further assumption is needed in models {\sf S15} to get a steeper oxygen
gradient.

A typical prescription is to assume an ``inside--out'' formation scenario for
the Galactic disc, namely that the infall timescale $\tau(r)$ is an increasing
function of $r$ (Chiosi \& Matteucci 1980; Lacey \& Fall 1983; Matteucci \& 
Fran\c{c}ois 1989). 
This can formally allow the model to predict a gradient close to the 
observed one (model {\sf S15b}, Fig.~\ref{gradOH}), 
but the required variation of the infall timescale from the centre to the 
outskirts of the Galaxy is too large to be physically acceptable: in model 
{\sf S15b} we need timescales
much longer than the age of the Universe ($\tau \sim 50$~Gyr at 
$r = 20$~kpc, see Table~\ref{modelstab}).
This would imply that the accretion process is still largely on-going;
most of the Galaxy mass on the outskirts would be still
accreting onto the disc nowadays. Such high accretion rates in spiral galaxies
at the present day are not observed in the X-rays (Bertin \& Lin 1996 
and references therein),
meaning that the accretion process should be mostly 
over by now.

We conclude that a simple Schmidt law is hardly suited at reproducing the 
observed metallicity gradients, at least within this kind of chemical models. 


\subsection{Models with Oort--type SF law {\sf (O10)}}

In the case of the SF law by Oort (Eq.~\ref{SFWyse} with $\kappa=1$),
model {\sf O10a} with a uniform infall timescale $\tau=3$~Gyr still
predicts too flat a metallicity gradient. 
To recover the observed gradient in the inside--out formation scenario,
as in the case of models {\sf S15} the required radial variation of $\tau$
becomes too extreme to be acceptable (model {\sf O10a}, 
Table~\ref{modelstab}). This kind of SF law also 
proves unable to reproduce the observed oxygen 
gradient, at least within the framework of these chemical models.


\subsection{Models with Dopita \& Ryder's SF law {\sf (DR)}}
\label{modelsDR}

Model {\sf DRa} is run with a uniform infall timescale {\mbox{$\tau$=3~Gyr}}
all over 
the disc. The predicted oxygen gradient is in good agreement with the data 
in the outer regions of the disc, while it flattens considerably in the 
inner regions (Fig.~\ref{gradOH}).

\begin{figure}[ht]
\centerline{\psfig{file=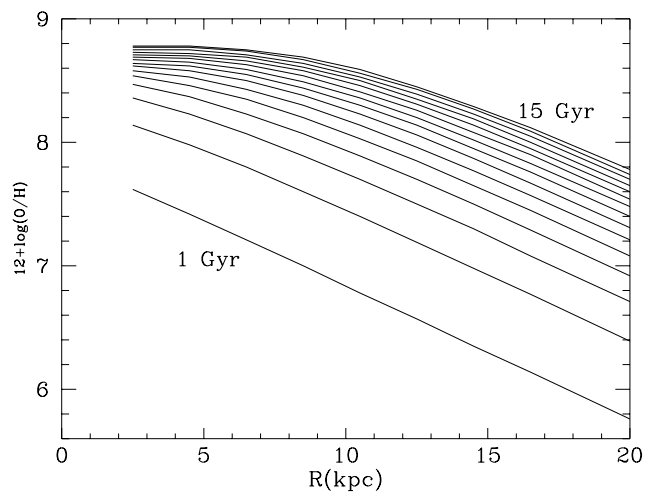,angle=-90,width=9truecm}}
\caption{Evolution of the oxygen gradient in model {\sf DRa}.
The gradient is plot at age steps of 1~Gyr, from 1 to 15 Gyrs.}
\label{drevolgradOH}
\end{figure}

It is worth commenting a bit further on this effect. A linear gradient in
[O/H]=12+log(O/H) implies an exponential decline of the metallicity $Z$. 
Therefore, a SF efficiency declining exponentially in radius, 
such as that by Dopita \& Ryder, looks appealing to account for 
the metallicity gradient.
In fact, simple closed models with such a SF law easily predict an 
exponential radial decrease of $Z$ (Edmunds \& Greenhow 1995).
But in more realistic models this prediction falls, and the metallicity
in the inner regions tends to settle onto some equilibrium value rather than
follow the espected exponential profile. As noticed by various authors 
(G\"otz \& K\"oppen 1992; Ferrini \etal 1994; Prantzos \& Aubert 1995; Ryder 
1995; Moll\'a \etal 1996, 1997), when infall is included and the IRA is 
relaxed, the model reaches a stage in which the 
metal enrichment provided by the ejecta of massive stars is significantly
diluted by the primordial infalling gas and by the gas expelled by long--lived,
low mass stars which formed at early times from relatively metal--poor gas. 
At this stage, metallicity 
abandons the one-to-one increase with SF processing and tends to settle on some
equilibrium value. This effect is best seen in the inner regions of the Galaxy,
where the SF process is more efficient: at the present age, most of the gas 
has been eaten up already and SF has so much reduced as to be 
compensated by the gas shed by the many low-mass stars which had formed in the
early, very active phases.~~As a consequence,~~in realistic models~the radial
metallicity profile tends to flatten in the inner regions. 
Fig.~\ref{drevolgradOH} shows how this effect develops in time: the gradient
is linear (in [O/H]) at early ages, while for 
$t \gsim 5$~Gyr the metal enrichment slows down in the central parts due to
the dilution effects of infall and of delayed return from low--mass stars.

Therefore, even in the case of an exponentially varying SF efficiency as
in models {\sf DR}, some further assumption is needed to steepen the abundance
profile in the inner regions. In the inside--out formation scenario, we 
take the infall process to be faster in the inner regions, while in the outer 
regions ($r \geq r_{\odot}$) we can just keep a uniform $\tau = 3$~Gyr (model 
{\sf DRb}, Fig.~\ref{gradOH}, dashed line).

This kind of SF law, combined with an inside--out formation scenario, can
reproduce the observed oxygen gradient within acceptable infall timescales 
over the whole disc.


\subsection{An intermediate case: models {\sf O15}}
\label{modelsO15}

Though with an Oort--type SF law the preferred empirical exponent is suggested
to be $\kappa=1$ (Kennicutt 1998), some authors adopt $\kappa=1.5$, which
may be still compatible with Kennicutt's observations provided the whole
disc model is considered altogether (Boissier \& Prantzos 1999; 
Prantzos \& Silk 1998).
For the sake of completeness, we briefly discuss this SF law as well
(Eq.~\ref{SFWyse} with $\kappa=1.5$).

The behaviour of models {\sf O15} is qualitatively intermediate between models
{\sf S15} or {\sf O10} and models {\sf DR}. Model {\sf O15a}, adopting 
a uniform infall timescale {\mbox{$\tau=3$~Gyr}} all over the disc, predicts
a metallicity gradient which is steeper than in models {\sf S15a}
or {\sf O10a}, but still shallower than the observed one, and shallower than
in model {\sf DRa}. The same flattening trend in the inner regions as in model
{\sf DRa} is also observed. With this SF law, model {\sf O15b} is tuned
to reproduce the observed gradient by means of a radial increase of the infall
timescale $\tau$, in the inside--out formation scenario (cfr.\
Table~\ref{modelstab}). At odds with model {\sf DRb}, here $\tau(r)$ needs \
to increase over the whole radial range, otherwise the predicted gradient 
is not steep enough in the outer regions; but, unlike models {\sf S15b} or 
{\sf O10b}, one does not need to resort to unplausible
timescales in the outskirts to recover the observed gradient.

Therefore, models {\sf O15} are acceptable toy--models (e.g.\
Prantzos \& Silk 1998). We will not discuss them any further though,
since we feel that an Oort--type SF law should be combined
with the empirical exponent $\kappa=1$ rather than 1.5 (Kennicutt 1998).
In all what follows, the reader might simply keep in mind that models {\sf O15}
would qualitatively display an intermediate behaviour between models 
{\sf S15}/{\sf O10} and models {\sf DR}.

\begin{figure}[ht]
\centerline{\psfig{file=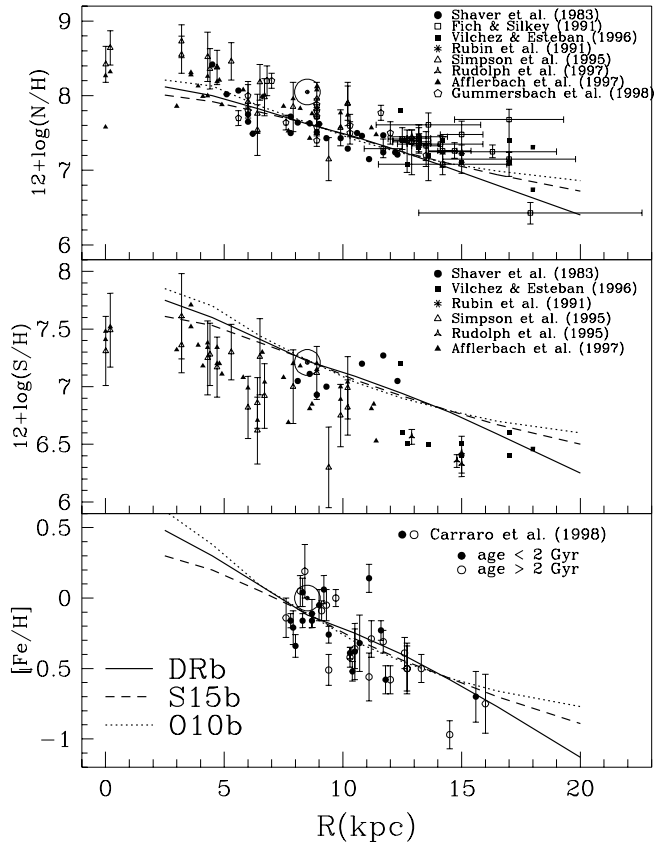,width=9truecm}}
\caption{Predicted nitrogen, sulfur and iron gradients from models indicated
in the legend on bottom left (see text) compared to observational data.}
\label{gradNSFeHfig}
\end{figure}


\subsection{Modelling the gradients of other elements.}

We have so far discussed the gradient of oxygen as the best tracer of the
overall metallicity. With respect to the radial gradients of the other elements
(\S~\ref{data}), all the models presented above which reproduce the observed 
oxygen gradient ({\sf S15b}, {\sf O10b}, {\sf DRb}) are also 
successful in predicting the other gradients 
(Fig.~\ref{gradNSFeHfig}). This simply means that the adopted nucleosynthetic
yields (from PCB98) are reliable. In fact, once a model is calibrated 
in $\nu$, $\zeta$ and $\tau$ so as to give the right oxygen abundance and 
gradient, the 
predicted behaviour of other elements merely depends on the relative abundance 
ratios implied by the yields.

Fig.~\ref{gradNSFeHfig} just shows some problem with the zero-point of the 
sulfur gradient (middle panel): the slope of the gradient is correct 
but the overall abundance is too high. This can be ascribed to the 
uncertainties in the yields of \Sulfur\ from SN\ae~II.
The mismatch between the predicted and observed sulfur abundance goes here 
in the opposite direction with respect to evidence from stars in the Solar 
Neighbourhood, which rather indicates that theoretical yields for \Sulfur\ 
are too low (PCB98). But the data on the \Sulfur\ abundance in 
low-metallicity local stars are quite few and old; further and improved 
observational data are needed to shed light on the problem of the yields of 
\Sulfur.

\bigskip \noindent
On the base of the constraint provided by the metallicity gradient, we can
consider {\sf DR} as viable models, while we tend to
exclude models {\sf S15} and {\sf O10}, and the corresponding
SF laws, since they would require unacceptably long accretion
timescales in the outskirts of the Galaxy. 
The situation may though change if radial inflows of gas are allowed for
(e.g.\ Lacey \& Fall 1985; G\"otz \& K\"oppen 1992;
Portinari \& Chiosi 1999). 

\begin{figure}[ht]
\centerline{\psfig{file=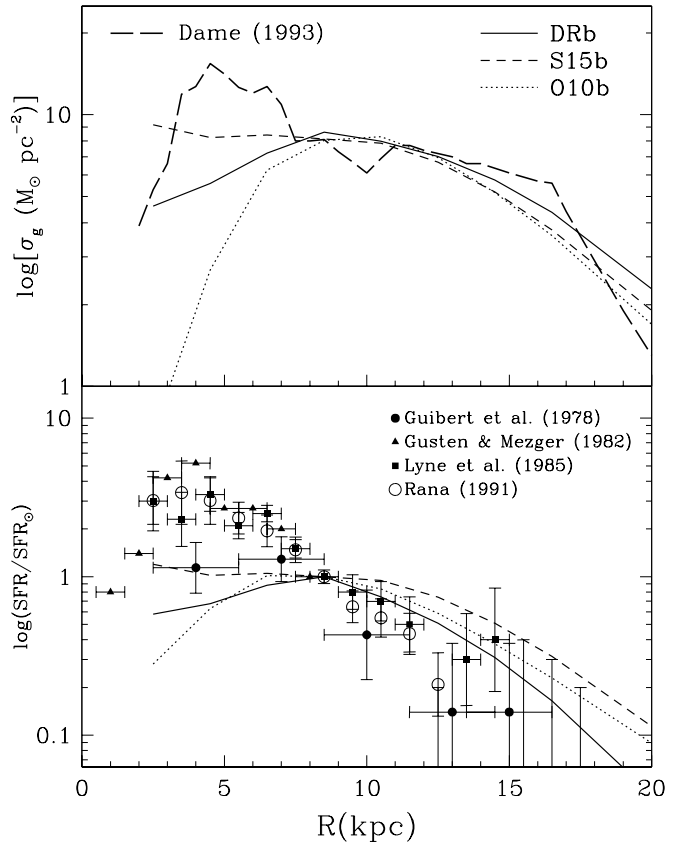,width=9truecm}}
\caption{Observed radial gas and (normalized) SFR distribution, together
with predictions from the models in the legend on top right.}
\label{radlGlSFRfig}
\end{figure}


\subsection{Modelling the gas and SF profiles}
\label{gasmodelling}

All of the models which are tuned, with more or less plausible model
parameters, to give successful predictions for the metallicity gradient 
({\sf S15b}, {\sf O10b}, {\sf DRb}) fall short at reproducing the observed 
radial profiles of gas density and SFR (Fig.~\ref{radlGlSFRfig}).
In particular, they cannot account for
the peak in the gas density corresponding to the molecular ring around 4 kpc,
and the corresponding peak in SF activity. Model {\sf S15} also seems to
predict too high a SFR outward of the Solar ring, in agreement with the
remark by Prantzos \& Silk (1998) that the present--day SFR along
the disc does not resemble any simple power law of the gas
density. This latter feature favours, in general, SF laws with a
radially varying efficiency (as in models {\sf DRb} and {\sf O10b}) 
with respect to simple Schmidt laws. 

\begin{figure}[ht]
\centerline{\psfig{file=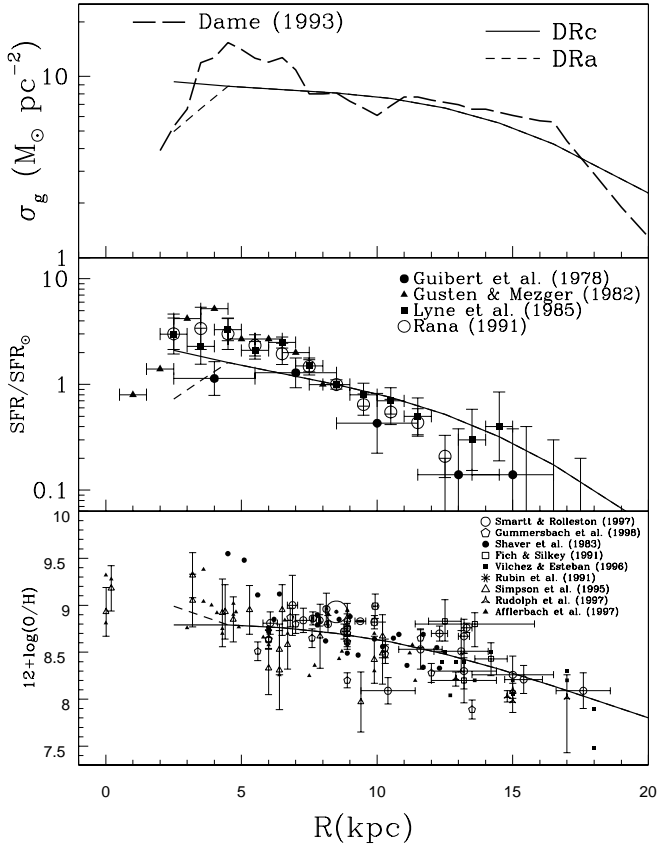,width=9truecm}}
\caption{Model {\sf DRc}, tuned so as to predict a maximum in the gas 
distribution around 4--5~kpc, still cannot reproduce the observed peak
(see text)}
\label{gastunedfig}
\end{figure}

In models {\sf S15b}, {\sf O10b}, {\sf DRb} the radial dependence of
the infall timescale was tuned so as to optimize the slope of the
predicted metallicity gradient (Fig.~\ref{gradOH}); these models tend to
yield a maximum in the gas distribution around $r_{\odot}$
(Fig.~\ref{radlGlSFRfig}). One can also adjust the infall timescale so
as to shift the maximum in gas density and SFR around 4--5~kpc, while keeping 
somewhat compatible with the observed metallicity gradient 
(e.g.\ model {\sf DRc} in Fig.~\ref{gastunedfig});
still, the sharp peak of the molecular ring is never recovered 
(see also Prantzos \& Silk 1998). 

{\it It seems to be impossible for these models to reproduce both the 
metallicity gradient and the radial gas distribution at the same time.}
This is easily interpreted in the framework of ``static'' chemical models, 
or models consisting of isolated rings, like ours. To produce a higher 
metallicity in the
inner regions, SF must have been  more effective there, and consequently the
available gas must have been more efficiently processed, consumed and locked
in long-lived stars or remnants. In the models, the present--day gas density 
decreases where the metallicity is higher, at odds with observations.

A possible way out, in the framework of simple static models, might be to 
assume that the IMF is skewed toward high masses in the inner regions 
(higher $\zeta$ or flatter power--law exponent). A higher percentage of 
massive stars provides in fact a more efficient chemical 
enrichment, while less gas remains locked in low--mass stars and the gas
density can keep high. But the issue of systematic variations in the IMF
is still very controversial, both on empirical and on theoretical grounds
(Scalo 1998a,b).

A less ``exotic'' possiblity than a strongly varying IMF along
the disc is that the formation of the molecular ring involves some 
kinematical or dynamical process, not accounted for in
simple static chemical models.
As is the case for rings in outer barred spirals, it is likely that 
the molecular ring in our Galaxy has formed because of gas accumulation in 
correspondance to the Lindblad resonances of a barred potential.
Models for the Galactic bar set its co--rotation between 2.5 and 3.5~kpc, 
and its Outer Lindblad Resonance between 4.5 and 6~kpc 
(e.g.\ Gerhard 1996 and references therein), which suggests a
correlation between the molecular ring and the dynamical effects of the bar. 
In order to reproduce these effects, chemical models should allow for
Bar--induced radial drifts of gas. Hereon the need to develop chemical models 
with radial flows of gas, which will be the subject of a second paper
(Portinari \& Chiosi 1999).


\subsection{The effects of a longer infall timescale}
\label{changetau}

All the models presented so far adopt an infall timescale of 3~Gyr
at the Solar ring, as the fixed point for the radial variation of
$\tau$ in the inside--out formation scenario. The accretion timescale for
open chemical models of the Solar Neighbourhood is usually fixed by the 
metallicity distribution of local G--dwarfs. 
For many years, the reference dataset was that by Pagel \& Patchett (1975),
later revised by Pagel (1989) and by Sommer--Larsen (1991a). This 
distribution was nicely fitted with infall timescales of 3--4~Gyr 
(e.g.\ Chiosi 1980; Matteucci \& Fran\c{c}ois 1989; Sommer--Larsen 1991a).

\begin{figure}[t]
\centerline{\psfig{file=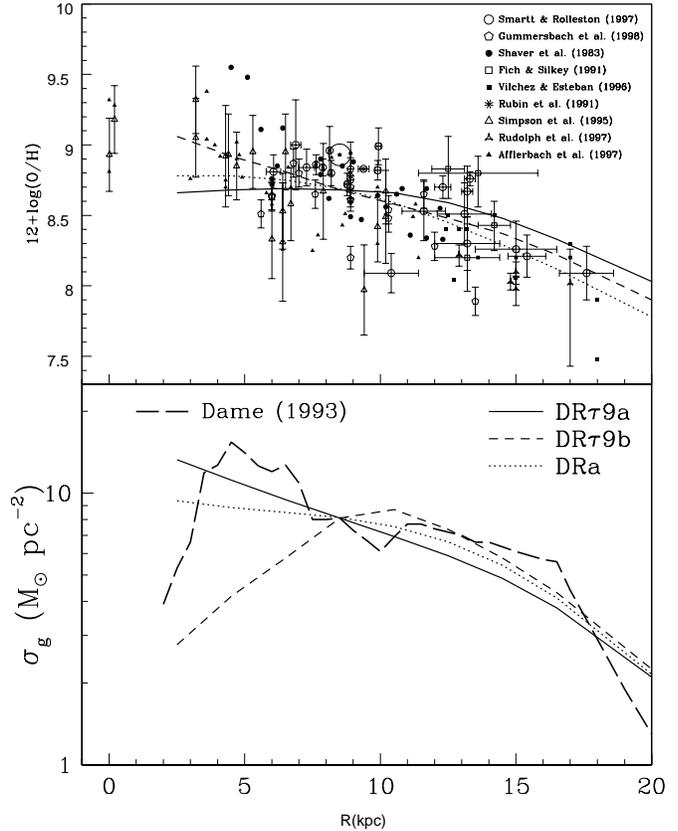,width=9truecm}}
\caption{Predicted oxygen gradient and gas distribution from model 
{\sf DR$\tau$9a} adopting a uniform infall timescale $\tau = 9$~Gyr 
(solid line), compared to model {\sf DRa} adopting $\tau = 3$~Gyr (dotted line)
and to the observational data. Model {\sf DR$\tau$9b}, adopting 
$\tau = 9$~Gyr at the Solar ring, is calibrated with $\tau$ increasing with 
radius so as to reproduce the observed gradient (see text). } 
\label{dopitatau9}
\end{figure}

The latest datasets by Wyse \& Gilmore (1995) and Rocha--Pinto \& Maciel 
(1996) display a narrower distribution of the stars in metallicity, 
with a prominent peak around [Fe/H]$\sim -0.2$, and therefore favour
much longer accretion timescales, up to 8--9~Gyr (Chiappini \etal
1997; PCB98). Such timescales may be uncomfortably long from the dynamical
point of view: in dynamical simulations the gas tends to settle onto the 
equatorial plane within the first 3--6 Gyrs (e.g.\ Burkert \etal 1992;
Sommer-Larsen 1991b; although Samland \etal 1997 suggest longer disc formation
timescales).

\begin{figure}[t]
\centerline{\psfig{file=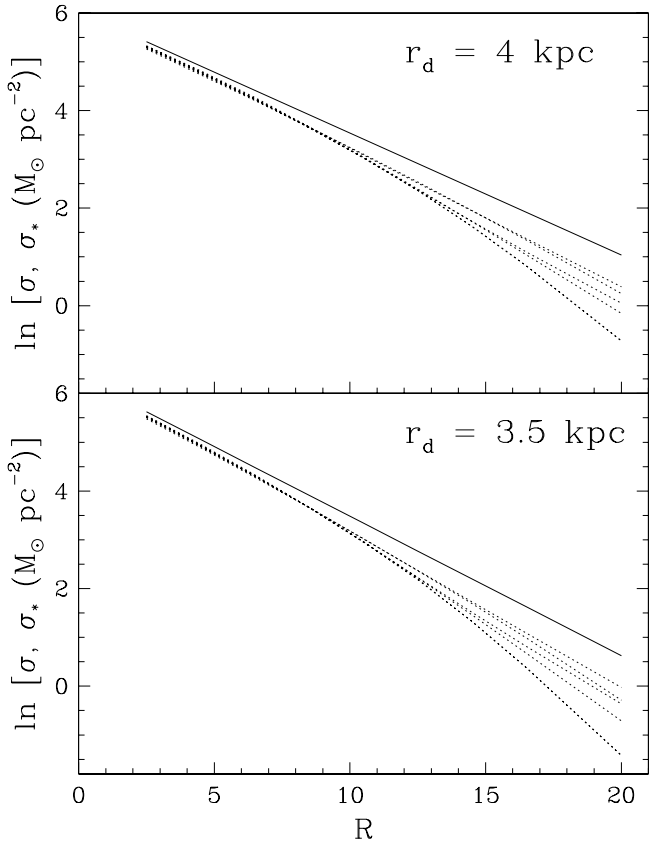,width=9truecm}}
\caption{{\it Top panel}: surface mass density profile (solid line) and
corresponding stellar density profile (dotted lines) for models with
$r_d=4$~kpc. {\it Bottom panel}: the same plot for models with $r_d=3.5$~kpc.}
\label{starsrdfig}
\end{figure}

Even more problematic here, if the adopted infall timescale is so long, it is 
hard for the chemical model to reproduce the observed metallicity gradient.
Let's consider, among the various SF laws studied here, the one
which intrinsically gives the steepest metallicity gradient, namely
Dopita \& {\mbox{Ryder's}} law. Model {\sf DR$\tau$9a} adopts a uniform infall
timescale $\tau=9$~Gyr, rather than 3~Gyr (Table~\ref{modelstab}).
The resulting
metallicity gradient is quite shallow (Fig.~\ref{dopitatau9}, solid line), 
much shallower than the gradient predicted by the analogous
model {\sf DRa} with $\tau=3$~Gyr (dotted line). In fact, when the infall 
timescale is very long, the dilution effect (\S~\ref{modelsDR}) is stronger 
and an equilibrium value for metallicity is reached over a larger
range of radii,
leading to an overall flat gradient. To reproduce the observed gradient within
the inside--out scenario (model {\sf DR$\tau$9b}, dashed
line) the assumed infall timescale for $r \gsim 10$~kpc needs to be of 
$\sim$50~Gyr (Table~\ref{modelstab}), implying too large infall rates at the 
present day. With other SF laws, things get even worse.

Radial inflows of gas in the disc can improve on this point, since they both
affect
the predicted G--dwarf distribution contributing to solve the G--dwarf problem,
and favour the formation of the metallicity gradient (Lacey \& Fall 1985; 
Clarke 1991; G\"otz \& K\"oppen 1992). This will be discussed in 
more detail in a forthcoming paper (Portinari \etal 1999, in preparation).

Finally, Fig.~\ref{dopitatau9} shows how it is still impossible to reproduce 
at the same time the metallicity gradient and the gas peak around 4~kpc,
also with models {\sf DR$\tau$9a,b}. Namely, the conclusions drawn in 
\S~\ref{gasmodelling} remain valid irrespectively of the adopted 
$\tau(r_{\odot})$.

\begin{figure}[t]
\centerline{\psfig{file=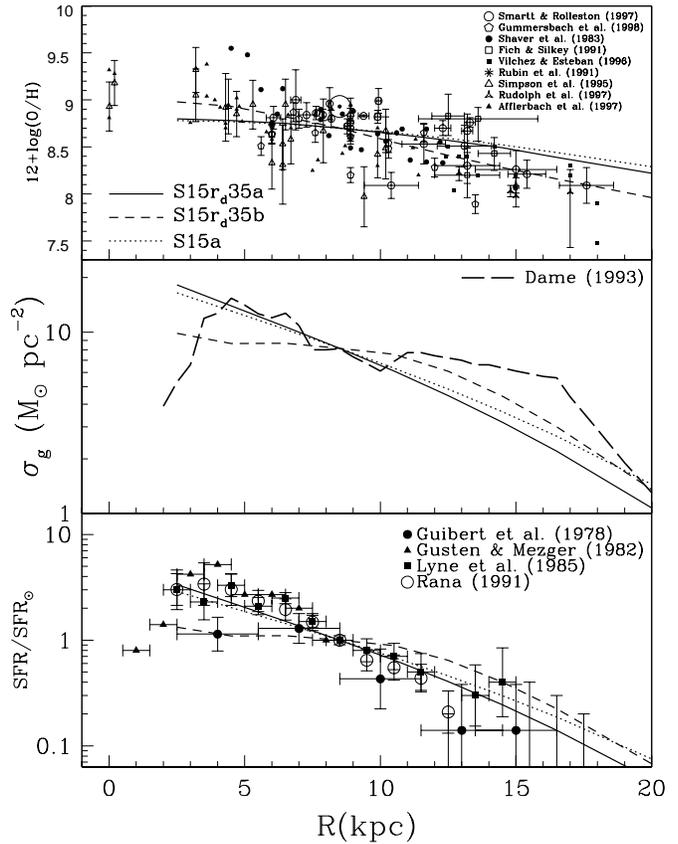,width=9truecm}}
\caption{Predicted oxygen gradient and gas and SFR profiles from model 
{\sf S15$r_d$35a} adopting a disc scale length $r_d = 3.5$~kpc
(solid line), compared to model {\sf S15a} adopting $r_d = 3.5$~kpc (dotted 
line) and to the observational data. Model {\sf S15$r_d$35b} is calibrated
with $\tau$ increasing with radius so as to reproduce the observed gradient 
(see text). } 
\label{s15rd35fig}
\end{figure}

\begin{figure}[ht]
\centerline{\psfig{file=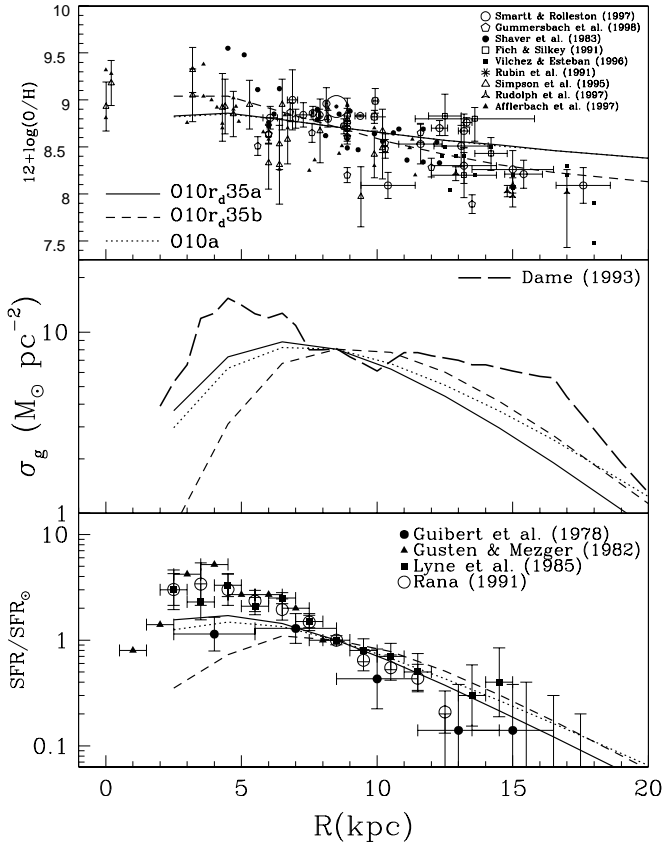,width=9truecm}}
\caption{Same as Fig.~\protect{\ref{s15rd35fig}}, but for models {\sf O10}.}
\label{ws10rd35fig}
\end{figure}

\begin{figure}[ht]
\centerline{\psfig{file=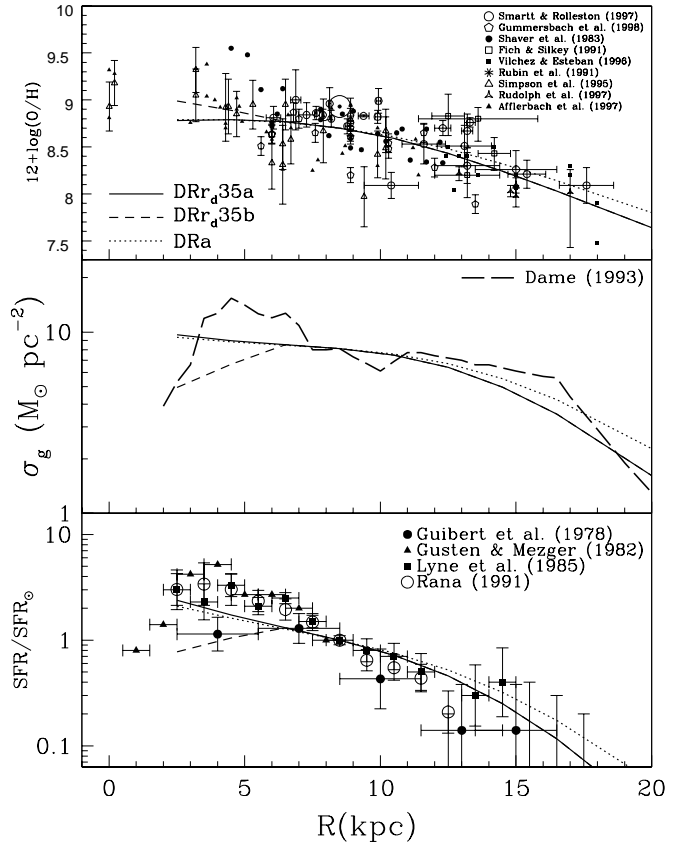,width=9truecm}}
\caption{Same as Fig.~\protect{\ref{s15rd35fig}} 
and~\protect{\ref{ws10rd35fig}}, but for models {\sf DR}.}
\label{drrd35fig}
\end{figure}


\subsection{The effects of a shorter disc scale--length}
\label{changerd}

All the models presented so far adopt an exponential scale length for mass
density $r_d=4$~kpc, quite typical of chemical evolution models 
(e.g.\ Sommer--Larsen 1991b, K\"oppen 1994, Rana 1991)
and consistent with early determination of the optical scale length of the 
Galactic disc (e.g.\ Binney \& Tremaine 1987 and references therein; 
Kuijken \& Gilmore 1989).
Recent observations, however, tend to suggest a shorter scale length for disc 
stars, $\lsim$3~kpc (Sackett 1997 and references therein, 
Vallenari \etal 1999 and references therein), comparable to that of other
Sbc galaxies. With $r_d=4$~kpc, our chemical models yield a final
(present-day) {\it stellar} density profile with scale length between 3 and
3.5~kpc (Fig.~\ref{starsrdfig}, top panel), still compatible but somewhat long 
with respect to the latest observations. Therefore, in this section we explore
models with a shorter mass density scale-length, $r_d=3.5$~kpc, corresponding 
to a final stellar density scale length around 2.5--3~kpc 
(Fig.~\ref{starsrdfig}, bottom panel).

Models {\sf S15$r_d$35a}, {\sf O10$r_d$35a} and {\sf DR$r_d$35a} correspond
exactly to models {\sf S15a}, {\sf O10a} and {\sf DRa} respectively, a part
from the adopted scale length (see Table~\ref{modelstab}).
The shorter scale length tends to favour the formation of the metallicity 
gradient and to predict a bit steeper decline of the gas density in the outer 
parts (compare solid to dotted lines in Fig.~\ref{s15rd35fig}
to~\ref{drrd35fig}). The effect is though
minor on models {\sf S15} and {\sf O10}, so that extreme infall timescales
still need to be invoked to reproduce the metallicity gradient in the 
inside--out scenario (models {\sf S15$r_d$35b} and {\sf O10$r_d$35b} in
Figs.~\ref{s15rd35fig} and~\ref{ws10rd35fig}; see also Table~\ref{modelstab}).
The effect of the steeper disc profile 
is more evident in models {\sf DR}, where the SF efficiency scales directly 
with surface density (\S~\ref{self-regSF}); nontheless, to reproduce the
observed metallicity gradient shorter infall timescales need to be assumed
in the inner regions, in accordance with the inside--out scenario.
The problem of reproducing the molecular ring is still present as well.

In the overall, adopting a shorter scale--length (within observational limits)
does not influence the bulk of the conclusions we drew in the previous 
sections.


\section{Summary and conclusion}
\label{conclusions}

From our analysis of chemical models for the Galactic disc with different
SF laws, it is evident that none of the SF laws considered is able, by 
itself, to reproduce the observed metallicity gradient throughout the whole 
extent of the disc. Some further ``dynamical'' assumption is needed in any 
case, such as an inside--out formation scenario (Larson 1976; Chiosi \& 
Matteucci 1980; Lacey \& Fall 1983; Matteucci \& Fran\c{c}ois 1989).

Even SF laws with an exponentially decreasing efficiency (e.g.\ Dopita \& 
Ryder's law), which in simple closed models would predict a logarithmic 
metallicity gradient like the observed one, display a more complex behaviour 
in detailed open models with no IRA. In particular, due to the late 
dilution of the enriched ejecta of massive
stars with infalling gas and returned gas from long-lived stars, the 
metallicity gradient tends to flatten in the inner regions with respect to the
outer slope.
To recover the observed slope, infall timescales need to be shorter in the
inner regions, to reduce the effects of dilution at late times. This shows 
how relaxing the IRA has non-negligible effects on chemical models, even 
in the case of oxygen which is produced on very short timescales and 
for which it is often assumed that neglection of finite stellar lifetimes 
is a viable approximation. Though this
might hold for the massive stars producing oxygen itself, the gas shed on 
delayed return from low-mass stars ultimately affects also the evolution of 
the oxygen abundance.

A Schmidt SF law with $\kappa=1.5$ or an Oort--type SF law with 
$\kappa=1.0$, though nicely fitting some empirical relations 
(Kennicutt 1998), are unable to account for the observed metallicity gradients
even in the inside--out picture, because the required radial variation of 
the infall timescale $\tau(r)$ implies accretion 
timescales in the outer regions much longer than a Hubble time.
These SF laws are therefore excluded, within the framework of simple ``static''
chemical models. These conclusions may change in the presence of radial flows 
of gas (Portinari \& Chiosi 1999).

We verified that these results are valid in general, irrespectively on the 
adopted
scale length of disc surface density (at least within observational estimates)
and on the adopted infall timescale at the Solar ring. If the latter is very
long, as maybe suggested from recent results on the ``G--dwarf problem'', even
SF laws of the Dopita \& Ryder type hardly reproduce the metallicity gradient
within the inside--out scenario.

Finally, with these ``static'' chemical models it turns out to be very 
difficult to reproduce at the same time both the metallicity gradient 
and the gas distribution, in particular the molecular ring (unless one resorts
to exotic possibilities like a varying IMF along the disc). Since the 
molecular ring is likely to be due to the dynamical influence of the 
Galactic bar, chemical models allowing for radial drifts and accumulation of 
gas might be needed to mimick these effects and reproduce the corrects gas 
distribution.
Our next step will be to improve the chemical model including radial
flows of gas (Portinari \& Chiosi 1999).



\acknowledgements{We thank Yuen K.\ Ng, Antonella Vallenari, Fulvio Buonomo 
and Giovanni Carraro for useful discussions on Galactic structure and dynamics,
and the referee whose comments improved the presentation of our paper.
L.P.\ acknowledges kind hospitality from the Nordita Institute in Copenhagen, 
from the Observatory of Helsinki and from Sissa/Isas in Trieste.
This study has been financed by the Italian Ministry for University and for
Scientific and Technological Research (MURST) through a PhD fellowship
and the contract ``Formazione ed evoluzione delle galassie'', n.~9802192401.}


{}

\end{document}